# Function + Action = Interaction

Ichiroh Kanaya, Mayuko Kanazawa, Masataka Imura

This article presents the mathematical background of general interactive systems. The first principle of designing a large system is to "divide and conquer", which implies that we could possibly reduce human error if we divided a large system in smaller subsystems. Interactive systems are, however, often composed of many subsystems that are "organically" connected to one another and thus difficult to divide. In other words, we cannot apply a framework of set theory to the programming of interactive systems. We can overcome this difficulty by applying a framework of category theory (Kleisli category) to the programming, but this requires highly abstract mathematics, which is not very popular. In this article we introduce the fundamental idea of category theory using only lambda calculus, and then demonstrate how it can be used in the practical design of an interactive system. Finally, we mention how this discussion relates to category theory.

## 1. Introduction

The Oxford English Dictionary (OED) defines the word "function" as: *1 an activity that is natural to or the purpose of a person or thing: 'bridges perform the function of providing access across water'; 'bodily functions'. 2 [Mathematics] a relation or expression involving one or more variables: 'the function (bx + c)'…*

As defined by the OED, the word function has (at least) a double meaning: activity (the first meaning) and relation (the second meaning). The relation is often referred to as a mapping in mathematics, which typically implies referential transparency.

Conversely, lambda calculus is identical to a mapping in mathematics, and also to a Turing Machine. This means that every program (computer code) can be represented by the lambda calculus [1].

However, there are often difficulties in regarding interactive systems as a mapping. Moggi tackled this problem and found a unique solution: he applied Kleisli category theory and regarded a function as a morph, which is a more abstract concept of an ordinary mapping [2].

Moggi discovered that these interactive systems cannot be divided into small lambda calculus expressions. He explained that this was due to the mismatch of *types* of input and output in the lambda calculus and suggested that this gap could be overcome by regarding an interactive system as a morph of a function to an action [3].

Moggi's theory uses highly abstract mathematics and is generally difficult for computer scientists to understand. However, Moggi's concept is understandable without a deep understanding of Kleisli's category theory.

In this paper we first interpret Moggi's discussion without using category theory and then explain how this theory can be applied to our interactive art called Polyphonic Jump! [4]. Finally, we provide rigid proof that our discussion follows Moggi's original proposal.

## 2. Interaction Equation

Let $x$ and $y$ be an input to and an output from a certain system, respectively. Generally $x$ and $y$ are not scalars. Hereafter, we assume that all functions are referentially transparent.

Interactive systems can be classified as one of four classes: A class 0 system outputs a constant value, while class 1 outputs a value that is a function of time. A class 2 system outputs a value that is a function of arbitrary inputs (including time). A class 3 system outputs a value that is a function of an arbitrary input and its internal status.

**Class 0:** output $y$ is constant, that is, $y = c$, where $c$ is a constant value.

**Class 1:** output $y$ is a function of time $t$. We denote this function as $f$, and call it the transfer function. Assume that all functions used in this paper are curried and left-associative. The equation for class 1 is

$$y = ft. \tag{1}$$

Class 2: output $y$ is a function of an arbitrary value $x$. Thus, the equation is

$$y = fx. \tag{2}$$

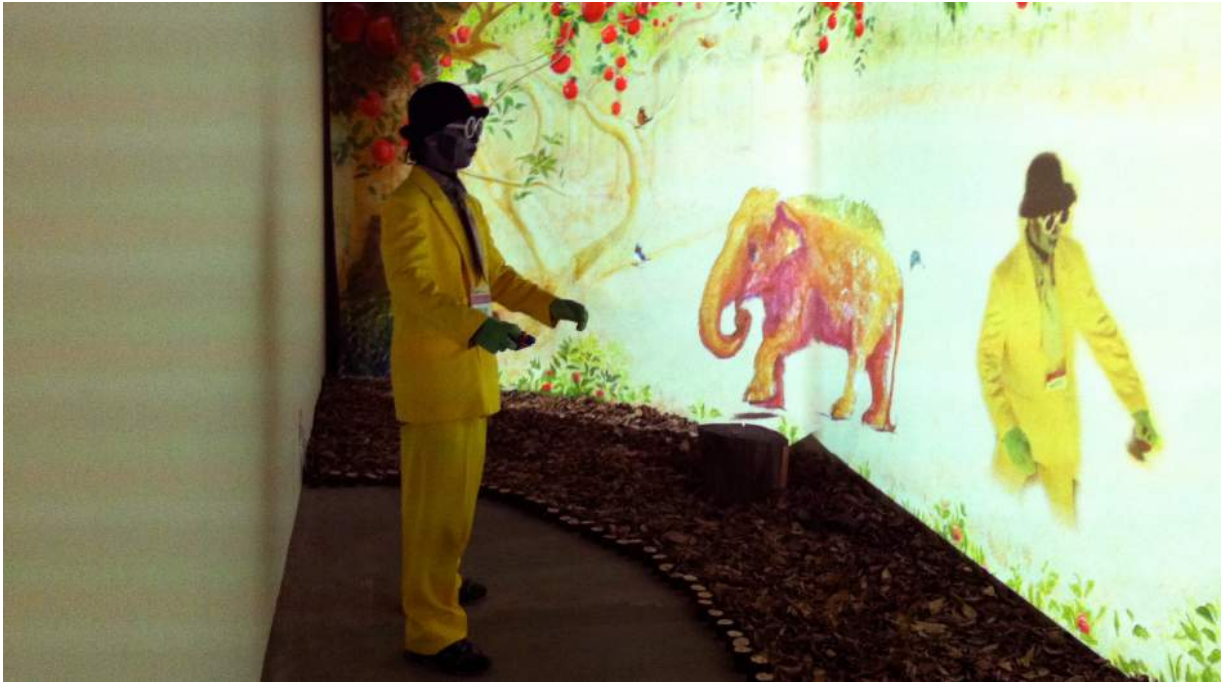

Figure 1. Polyphonic Jump!

Class 3: output $y$ is a function of an arbitrary input $x$ and an internal status $s$. If we allowed referential opacity of function $f$, we would obtain the following equation:

$$y = f_! \; s \; x, \quad (3)$$

where function $f_!$ changes its behavior based on $s$, and rewrites the value of $s$. Because rewriting any variables is not allowed in this discussion, we forget about this disruptive function $f_!$.

One well-known method for retaining referential transparency is placing the internal status outside the box. For example,

$$[y, t] = f'[x, s] \quad (4)$$

is a referentially transparent equation. Here function $f'$ returns a pair of output $y$ and a new internal status $t$. To match the types of input and output, the argument is also a pair.

Assume we have function inject given by

$$\mathit{inject} \; x := \backslash s \, . \, [x, s] \quad (5)$$

where $\backslash$ denotes lambda. This function inject abstracts the internal status $s$, and thus we can call inject $x$ as an input with context.

Because we wish to apply transfer function $f$ to input $x$, output $y$ should be

$$y = \mathit{inject} \; (f \; x). \quad (6)$$

Although Equation (6) is perfectly correct, it is not practical, because output $y$ is with context, but input $x$ is without context. A practical transfer function, say $F$, would be

$$y = F \; (\mathit{inject} \; x). \quad (7)$$

Let us extract transfer function $f$ as a parameter of function $F$ as

$$F = \mathit{bind} \; f \quad (8)$$

to obtain

$$y = \mathit{bind} \; f \; (\mathit{inject} \; x). \quad (9)$$

Equation (9) of a class 3 system corresponds with equation (2) of a class 2 system.

Next we introduce several symbols to facilitate human readability. The dagger symbol denotes the injection operator and is given by

$$x^\dagger := \mathit{inject} \; x. \quad (10)$$

The # symbol denotes the binding operator and is given by

$$f \; \# \; m := \mathit{bind} \; f \; m. \quad (11)$$

Equation (9) can be simplified by these operators as follows:

$$y = f \; \# \; x^\dagger \quad (12)$$

The binding operator is defined as

$$f \; \# \; m := \backslash s \, . \, f \; x \; s' \; \textbf{where} \; [x, s'] = m \; s \, , \quad (13)$$

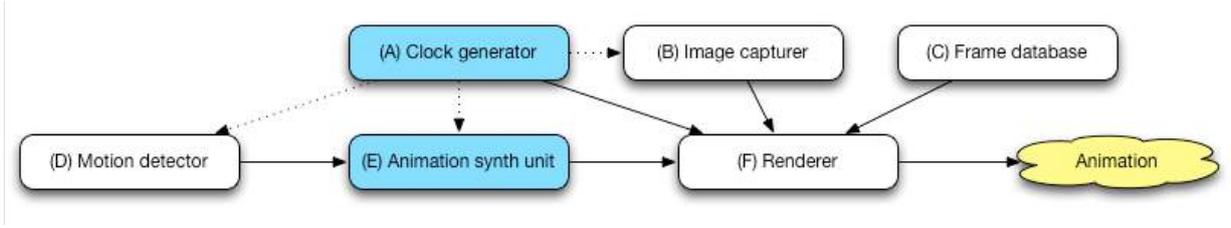

Figure 2. System structure of Polyphonic Jump!

where the keyword **where** declares local variables. Because almost all practical programming languages provide syntax for declaring local variables, we follow popular programming-language style and use let instead of where:

$$f \# m := \backslash s \,.\, \textbf{let}\ [x, s'] = m\ s\ \textbf{in}\ f\ x\ s' \qquad (14)$$

where **let** … **in** … is defined as

$$\textbf{let}\ a = b\ \textbf{in}\ c := (\backslash a \,.\, c)\ b. \qquad (15)$$

## 3. Composition of Transfer Functions

Assume that transfer function $f$ is the composition of two different transfer functions $g$ and $h$; that is,

$$f = h \bullet g, \qquad (16)$$

where

$$b \bullet a := \backslash z \,.\, b\ (a\ z). \qquad (17)$$

Let $x$ be the input to the system, and $m$ be a contextual version of $x$. Then m is given by

$$m := x^\dagger. \qquad (18)$$

Furthermore, let

$$n := g \# m \qquad (19)$$

where

$$n = \backslash s \,.\, \textbf{let}\ [x, s'] = m\ s\ \textbf{in}\ g\ x\ s'. \qquad (20)$$

Now we can expand $h \# n$ as

$$h \# n = \backslash t \,.\, \textbf{let}\ [x, s'] = m\ t, [y, t'] = g\ x\ s'$$
$$\textbf{in}\ h\ y\ t', \qquad (21)$$

which leads to

$$h \# g \# {}^*x = \backslash s \,.\, \textbf{let}\ [x, s'] = {}^*x\ s, [\_, s''] = g\ x\ s'$$
$$\textbf{in}\ (h \bullet g)\ x\ s''. \qquad (22)$$

As seen above, transfer functions can be combined using a binding operator. Composition through a binding operator keeps the context as shown in equation (22), and also maintains the order of evaluation of the functions because the operator follows equation (17).

Now we use another operator $\$$ denoting quick composition of transfer functions. We can think of applying non-contextual function $f_{NC}$ to contextual m as

$$f_{NC}\ \$\ m := (f_{NC}\ x)^\dagger\ \textbf{where}\ \backslash s \,.\, [x, s'] = m\ s. \qquad (23)$$

Operator $\$$ gives context to function $f_{NC}$, and is known as a *functor* as discussed later.

## 4. An Example: Polyphonic Jump!

Polyphonic Jump! is a system that allows humans to be immersed in a fantasy world in which many creatures create a polyphonic chorus. The audience stands in front of a huge canvas on which a picture of a forest has been painted in oils, and individuals jump to interact with oil-painted animals on the canvas as if they were also on the canvas. These individuals feel as though they are actually in a picture book [4].

For seamless integration of the physical painting, which presents true reality and computer-generated animation that moves dynamically and interacts with the audience, we have incorporated real-time 3D modeling and projection technology in this artwork (see Figure 1).

As shown in Figure 2, Polyphonic Jump! has the following subunits: (A) clock generator, (B) image capturing unit, (C) animation frame database, (D) motion sensor, (E) animation generator, and (F) renderer. The authors use white for trivial referentially transparent units, and blue for non-trivial referentially transparent units. Arrows show the flow of information.

**(A)** The clock generator synchronizes all units by controlling the renderer (F).

**(B)** The image capturing unit captures a figure in the audience.

**(C)** The animation frame database retrieves each frame of animations.

**(D)** The motion sensor returns True if a member of the audience is jumping, otherwise False.

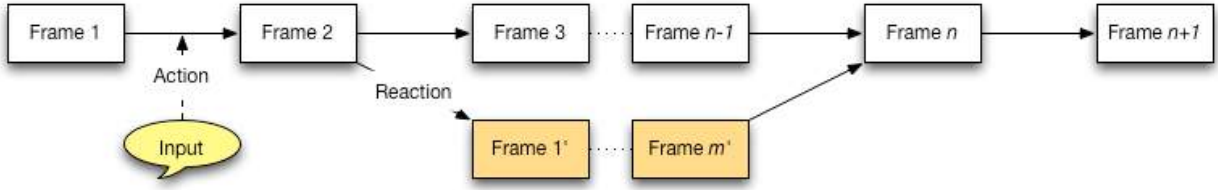

Figure 3. Example of animation sequence of Polyphonic Jump!

**(E)** The animation generator, referring to the motion sensor (D), generates frame information in XML format based on current time. Animation in this art work is complex because multiple sequences run at different timings/speeds.

**(F)** The renderer renders a frame based on the XML information given by the animation generator (E) and images from the animation frame database (C).

Units (B), (C), (D), and (F) are trivially referentially transparent, because unit (B) is a function that takes a time value and returns an image, unit (C) is a function that takes a query and returns images, unit (D) is a function that returns the audience's motion, and unit (F) is a function that takes the frame information and returns a computer-graphics image.

Unit (A) returns a time-variant value, however, it is still referentially transparent when considering that it always returns an "evaluate the current timing" action.

Conceptually unit (E) has its own internal status, because it runs a pre-defined animation sequence (normal status), and starts a new animation when the motion sensor triggers the unit (triggered status). After a certain time, unit (E) returns to its normal status.

The actual unit (E) was designed to be completely referentially transparent. The internal status is given, and proceeds through the unit as an action (lambda calculus). This action is eventually evaluated in unit (F) once rendering has started.

Polyphonic Jump! assigns time (action to evaluate the current time) to variable x in equation (5), and the status of the animation generator as context s.

## 5. Note on Monad of Category Theory

We define a category $C$ with objects $A, B, \ldots$ and a morph $\phi$. Objects are monoids, including the set of integers, list of scalars, and tree of scalars. Morph $\phi$ can be a function length that returns the length of a list.

If we have an identity projection $id_C$ and a functor $T$ from category $C$ to $C$, the following natural transforms $\eta$ and $\mu$ follow:

$$\eta: id_C \rightarrow T, \quad (24)$$

$$\mu: T^2 \rightarrow T. \quad (25)$$

Moreover, if transforms $\eta, \mu$ are commutative with functor $T$, i.e., $\eta_{TA} = T\eta_A$ and $T\mu_A = \mu T_A$, a triple $[T, \eta, \mu]$ is called a monad in category theory [2].

Kleisli introduced operator * instead of the $\mu$ of category theory, and called triple $[T, \eta, *]$ a Kleisli triple. Operator * follows these equations:

$$(\eta_A)^* = id_{TA}, \quad (26)$$

$$f^* \bullet \eta_A = f, \quad (27)$$

$$g^* \bullet f^* = (g^* \bullet f)^*, \quad (28)$$

where projection $f$ projects $A$ to $TB$ and another projection $g$ exists. Figure 4 illustrates the relationship among functor $T$, the natural transform $\eta$, and operator *.

Kleisli's triple is identical to our triple $[\$, †, \#]$, which is called a Monad in Programming. For example, the triple *[fmap, return, >>=]* in the programming language Haskell is identical to Kleisli's triple.

## 6. Concluding Remarks

In this paper we presented a strict mathematical framework for interactive systems. A difficulty in describing such interactive systems relates to dividing such systems into subsystems owing to the organic connection of every part of the system. Global variables, hidden contexts, and non referentially transparent functions are examples of this difficulty in programming [5].

Referential transparency is a popular concept among mathematicians for reducing complexity. We can regard a function as a projection of values if the function is referentially transparent. The domain and co-domain of a function are monoids if they have an identity projection. This means that such a projection can intuitively be divided into composite projections, thereby reducing the complexity for programmers. For this reason, some domain specific languages for scientific computing support referential transparency [6, 7].

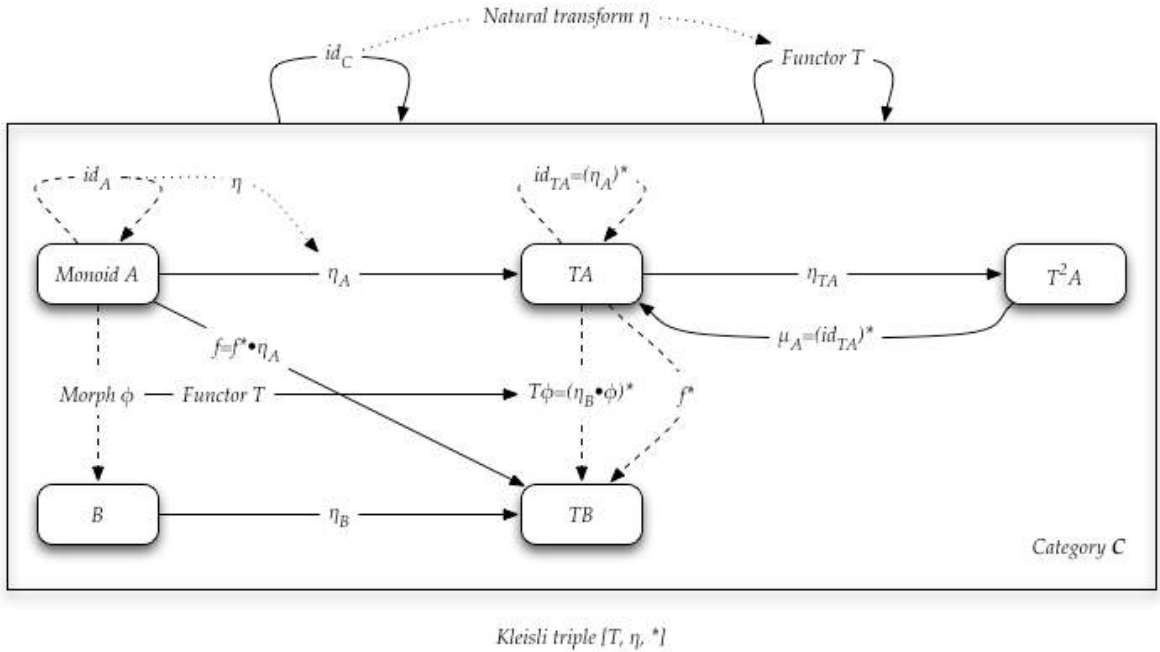

Figure 4. Relationship of operators *T, η, \** of Kleisli Triple *[T, η, \*]*

Algorithms for interactive systems must consider both the input from users and output to users, and thus they cannot be discussed simply as purely mathematical mappings. For example, composition of monodis is well studied and can be applied to scientific computing, however, it cannot be applied to interactive systems.

Interactive systems are, however, projections (morphs) in terms of the Kleisli category. The Kleisli triple is identical to the monad of programming.

This paper showed that interactive systems can be described as a composition of subsystems without using highly abstract mathematics. It also illustrated the concrete example of Polyphonic Jump! and showed how our discussion corresponds with traditional category theory.

Referential transparency is not the only way to divide interactive systems into subsystems. The monad of programming can spatially divide a system, while the continuation of programming can temporally divide a system. Unfortunately continuation is known to disrupt referential transparency; however, we can still hope for the existence of a more abstract mechanism that treats referential transparency and continuation equally.